\documentclass[9pt,twocolumn,twoside]{osajnl}
\usepackage{exscale}
\usepackage{siunitx}
\usepackage{relsize}
\usepackage{graphicx,subfigure,amssymb,dcolumn,bm,amsmath}
\usepackage{color} 
\journal{ol}
\setboolean{shortarticle}{true}
\title{Stable higher-charge vortex solitons in the cubic-quintic medium with a ring potential}
\author[1*]{Liangwei Dong}
\author[2]{Mingjing Fan}
\author[3, 4]{Boris A. Malomed}

\affil[1]{Department of Physics, Zhejiang University of Science and Technology, Hangzhou, China, 310023}
\affil[2]{Department of Physics, Shaanxi University of Science and Technology, Xi'an, China, 710021}
\affil[3]{Department of Physical Electronics, School of Electrical Engineering, Faculty of Engineering, Tel Aviv University, Tel Aviv 69978, Israel}
\affil[4]{Instituto de Alta Investigacion, Universidad de Tarapaca, Casilla 7D, Arica, Chile}

\affil[*]{Corresponding author: dlw\_0@163.com}

\ociscodes {(190.0190) Nonlinear optics; (190.6135) Spatial solitons; (070.7345) Wave propagation.}

\begin{abstract}
We put forward a model for trapping stable optical vortex solitons (VSs) with high topological charges $m$. The cubic-quintic nonlinear medium with an imprinted ring-shaped modulation of the refractive index is shown to support two branches of VSs, which are controlled by the radius, width and depth of the modulation profile. While the lower-branch VSs are unstable in their nearly whole existence domain, the upper branch is completely stable. Vortex solitons with $m\leq 12$ obey the anti-Vakhitov-Kolokolov stability criterion. The results suggest possibilities for the creation of stable narrow optical VSs with a low power, carrying higher vorticities.
\end{abstract}
\setboolean{displaycopyright}{true}

\begin{document}

\maketitle

%

Optical vortices are a source of numerous phenomena in physics of light,
displaying deep similarities to quantized vortices in superfluids and
Bose-Einstein condensates (BECs) \cite{Prog2005}. In addition to the broad
phenomenology, the power profile of vortices and orbital angular momentum
carried by them offer various applications, such as optical trapping,
tweezers, data processing, quantum communications, etc. \cite{shen2019optical}.

Self-focusing nonlinearity usually leads to azimuthal instability and
splitting of vortex-carrying beams. Competing nonlinearities, which suppress
the self-focusing-driven collapse, in some cases may also stabilize vortex
solitons (VSs) against the spontaneous splitting \cite{MALOMED2019108}.
Known examples include quadratic-cubic \cite{PhysRevE.66.016613},
cubic-quintic (CQ) \cite{Quiroga-Teixeiro97,PhysRevE.64.057601,Pego,michinel2004square} and
quintic-septimal \cite{PhysRevA.90.063835} optical media, with lower-order
focusing and higher-order defocusing nonlinear terms. In particular, some
optical materials can be accurately approximated by the CQ model \cite{Lawrence:98,Cid}. Competing nonlinearities also occurs in plasmas \cite{zakharov1971behavior} and BEC \cite{PhysRevLett.78.1215,Petrov}.

Relevant alternatives for trapping stable VSs are provided by confinement,
with spatial modulation of the refractive index (RI) inducing an effective
trapping potential \cite{Mazilu,RevModPhys.83.247, MALOMED2019108,BorisBook2022,PhysRevLett.130.157203}. Examples include graded-index optical
fibers \cite{Raghavan2000377}, nonlinear photonic crystals with defects \cite{2004OExpr..12..817F}, as well as linear and nonlinear optical lattices \cite{dror2015solitons}. The trapped VSs follow the structure of the underlying
RI modulation \cite{RevModPhys.83.247,MALOMED2019108,BorisBook2022}. A similar potential profile maintains stable VSs in BEC \cite{PhysRevLett.84.806}.

Experimentally, robust vortex modes were observed in saturable \cite{PhysRevA.93.013840} and CQ media \cite{wu2013cubic,Reyna:16}. Theoretically, in most cases stable VSs were predicted with topological charge up to $m=2$ \cite{Yang:03, MALOMED2019108, BorisBook2022}, with a few
exceptions \cite{PhysRevLett.95.123902,dong2011higher,PhysRevLett.115.193902,Ecnu,Raymond,PhysRevLett.129.123903,LIU2023113422}. In particular, in free space the 2D nonlinear Schr\"{o}dinger equation (NLSE) with the CQ nonlinearity produces relatively broad stability areas for $m=1$ and $2$, and extremely narrow ones for $m\geq 3$ \cite{Pego,michinel2004square}, while the interplay of the  harmonic-oscillator trapping potential and cubic self-focusing makes it possible to stabilize
solely VSs with $m=1$ \cite{Mazilu}. In nonlocal media, stable higher-charge VSs can exist as a component of a vector soliton, guided by the RI profile induced by a copropagating fundamental-mode component \cite{zhang2020instability,zhang2022stabilization}. Discrete VSs with charge exceeding $2$ can propagate stably in conventional \cite{dong2011higher}, parity-time-symmetric \cite{PhysRevLett.115.193902}, twisted \cite{PhysRevLett.129.123903}, and semidiscrete \cite{Raymond2} waveguide arrays.

The present work aims to demonstrate that stable vortex
modes with higher charges can be supported by a combination of the CQ
nonlinearity and an effective ring-shaped guiding potential. We find that
VSs with different charges bifurcate, as lower- and upper-branch states,
from the corresponding linear eigenmodes supported by the same potential.
The upper-branch vortices with $m$ up to $12$ (at least) are {\em completely
stable} in their entire existence domain. The waveguide with the required
structure can be designed by means of the well-known technique used for the
fabrication of multilayer optical fibers \cite{Payne}.


%

We consider the propagation of a light beam in the bulk CQ medium with the
imprinted RI modulation, governed by the 2D NLSE, written in the scaled
form:
\begin{equation}
\centering i\frac{\partial \Psi }{\partial z}=\left[ -\left( \frac{\partial
^{2}}{\partial x^{2}}+\frac{\partial ^{2}}{\partial y^{2}}\right)
-pV(r)-|\Psi |^{2}+|\Psi |^{4}\right] \Psi ,  \label{Eq1}
\end{equation}%
Here $z$ and $(x,y)$ are the propagation distance and transverse
coordinates, while both the cubic and quintic coefficients are scaled to be $1$, without the loss of generality. As ring-shaped trapping
potential is favorable for the formation of VSs, we adopt the RI
modulation profile,
\begin{equation}
V(r)=\exp \left[ -(r-r_{0})^{2}/d^{2}\right] ,  \label{V}
\end{equation}
where $r=\sqrt{x^{2}+y^{2}}$, $r_{0}$, $d$ and $p$ being the radius, width,
and depth of the ring potential, respectively.

Stationary solutions of Eq.~(\ref{Eq1}) with propagation constant $b$ are
looked for as $\Psi (x,y,z)=\psi (x,y)\exp (ibz)\equiv \left[\psi_{r}(x,y)+i\psi_{i}(x,y)\right] \exp (ibz)$, where $\psi _{r}$ and $\psi
_{i}$ are the real and imaginary parts of the stationary mode, whose phase
is $\phi =\arctan \left( \psi _{i}/\psi _{r}\right) $. The substitution of
this ansatz in Eq.~(\ref{Eq1}) yields a stationary equation,
\begin{equation}
\centering\frac{\partial ^{2}\psi }{\partial x^{2}}+\frac{\partial ^{2}\psi
}{\partial y^{2}}+pV(r)\psi -b\psi +|\psi |^{2}\psi -|\psi |^{4}\psi =0,
\label{Eq2}
\end{equation}%
which was solved by dint of the relaxation or Newton-conjugate-gradient
method \cite{Book2}. Parameters of the soliton families are $p$, $r_{0}$, $d$%
, and $b$. By means of scaling, we fix $r_{0}=2\pi $ and, as a typical
relevant value of the ring's width in Eq. (\ref{V}), take $d=\sqrt{6}$,
varying $p$ and $b$ to produce vortex-soliton families. Equation (\ref{Eq1})
conserves net power (norm) $P$, Hamiltonian $H$, and angular momentum $M$:
\begin{equation}
\begin{aligned} &P=\int \int |\psi |^{2}\text{d}x\text{d}y, \label{Eq3} \\
&H=\mathlarger{\int}\mathlarger{\int} \left[ \left\vert \frac{\partial \psi
}{\partial x}\right\vert^{2}+\left\vert \frac{\partial \psi }{\partial
y}\right\vert ^{2}+pV|\psi |^{2}-\frac{1}{2}|\psi |^{4}+\frac{1}{3}|\psi
|^{6}\right] \text{d}x\text{d}y, \\ &M=i\mathlarger{\int}\mathlarger{\int}
\left[ \psi ^{\ast }\left( y\frac{\partial }{\partial x}-x\frac{ \partial
}{\partial y}\right) \psi \right] \text{d}x\text{d}y\equiv i\int \int \psi
^{\ast }\frac{\partial }{\partial \theta }\psi \text{d}x\text{d}y,
\end{aligned}
\end{equation}%
with $\ast $ standing for the complex conjugate. Vortex solitons with
topological charge $m$ have the form of $\Psi (x,y,z)=|\psi (x,y)|\exp
(im\theta +ibz)$, which gives rise to relation $M=mP$.

The stability of the VSs was investigated by taking perturbed solutions to
Eq.~(\ref{Eq1}) as $\Psi (x,y,z)=[\psi (x,y)+u(x,y)\exp (\delta z)+v^{\ast
}(x,y)\exp (\delta ^{\ast }z)]\exp (ibz)$, where $u$ and $v$ are
infinitesimal perturbations and $\delta $ is the instability growth rate.
Linearization of Eq.~(\ref{Eq1}) around $\psi $ yields an eigenvalue
problem:
\begin{equation}
i%
\begin{bmatrix}
\mathcal{M}_{1} & \mathcal{M}_{2} \\
-\mathcal{M}_{2}^{\ast } & -\mathcal{M}_{1}^{\ast }%
\end{bmatrix}%
\begin{bmatrix}
u \\
v%
\end{bmatrix}%
=\delta
\begin{bmatrix}
u \\
v%
\end{bmatrix}%
,  \label{Eq4}
\end{equation}%
%
where ${\mathcal{M}_{1}}=\partial _{x}^{2}+\partial
_{y}^{2}+pV-b+2\left\vert \psi \right\vert ^{2}-3|\psi |^{4}$ and ${\mathcal{%
M}_{2}}=\psi ^{2}(1-2|\psi |^{2})$. Equations (\ref{Eq4}) were solved by
means of the Fourier collocation method \cite{Book2}. Solitons may be stable
if all eigenvalues $\delta $ are imaginary.



\begin{figure}[tbph]
\centering
\includegraphics[width=0.48\textwidth]{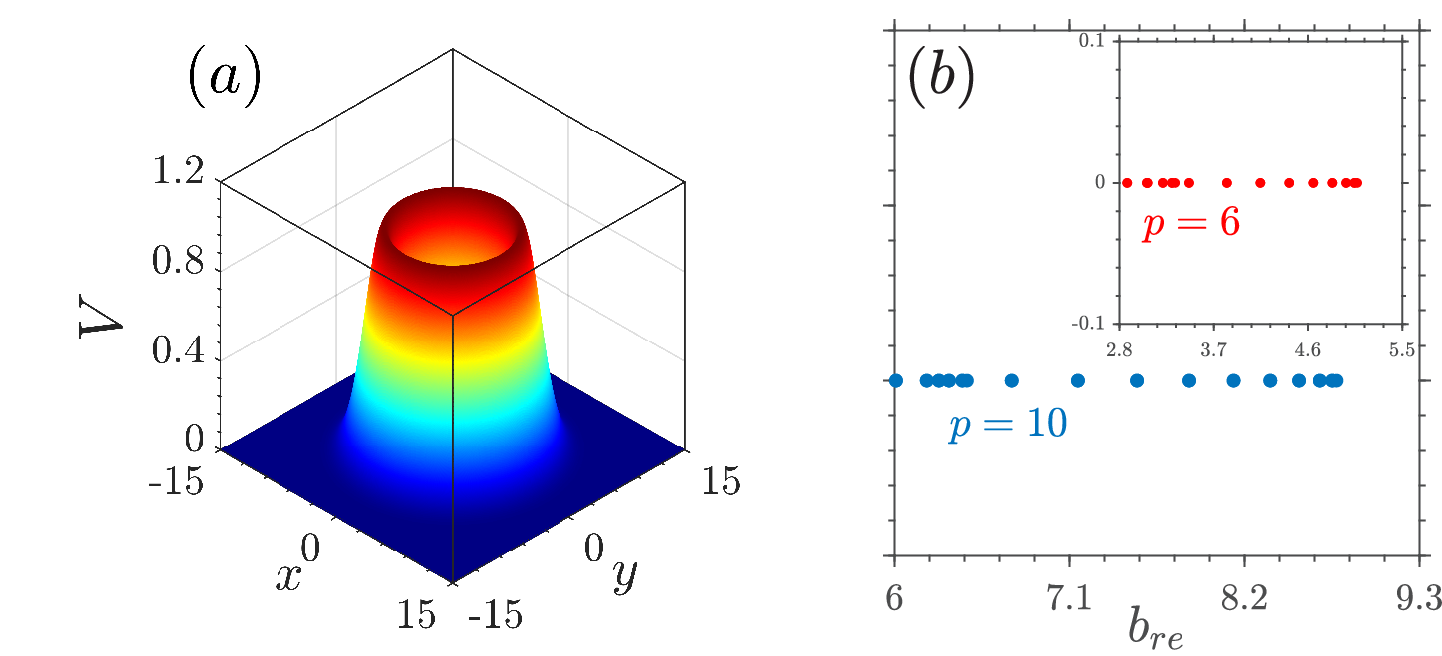}\vskip-0pc
\caption{(a) The ring-shaped potential (\protect\ref{V}). (b) Spectra of
real eigenvalues $b_{\text{re}}$ of the linearized equation (\protect\ref%
{Eq2}) for different values of $p$.}
\label{fig1}
\end{figure}

First, it is relevant to produce the dispersion relation for the linear
system with the ring-shaped potential (\ref{V}). An example of the potential
and respective set of discrete eigenvalues, produced by the numerical
solution of the linear version of Eq.~(\ref{Eq2}), are shown in Figs.~\ref%
{fig1}(a) and (b), where the growth of the potential's depth $p$ results in
a shift of the spectrum to the right.

In the framework of the nonlinear equation (\ref{Eq2}), fundamental and
dipole solitons bifurcate, respectively, from the linear ground-state
eigenmode, $\psi _{0}$, and the degenerate pair of mutually perpendicular
dipole ones representing the lowest excited states, $\psi _{1,1}$ and $\psi
_{1,2}$. Vortex solitons are produced by their superposition as $\psi
_{m=\pm 1}=\psi _{1,1}\pm i\psi _{1,2}$. Similarly, VSs with topological
charge $m$ bifurcate from the superposition of the $m$-th pair of linear
eigenmodes as $\psi _{\pm m}=\psi _{m,1}\pm i\psi _{m,2}$.

\begin{figure}[tbph]
\centering
\includegraphics[width=0.48\textwidth]{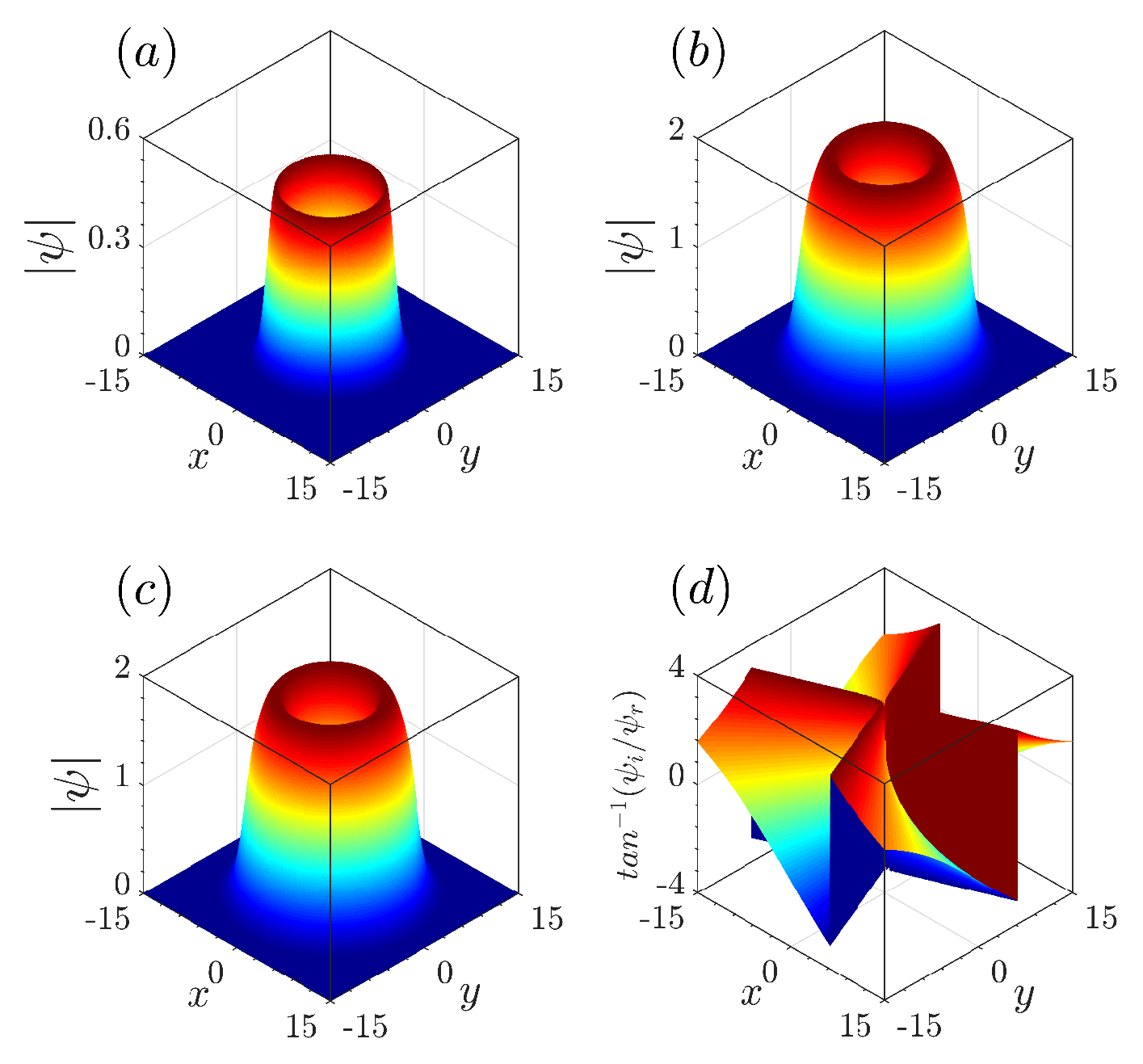}\vskip-0pc
\caption{Profiles of stable VSs marked in Fig. \protect\ref{fig3}(a), with $%
m=2$ (a,b) and $m=4$ (c), supported by the ring potential (\protect\ref{V})
with $p=10$ and $r_{0}=2\protect\pi $. (d) The phase pattern corresponding
to (c). The propagation constant is $b=8.8$ (the lower branch) in (a), and $%
1.5$ (the upper branch) in (b-d).}
\label{fig2}
\end{figure}

To build VSs, it is relevant to address the setting with the potential's depth $p$ of the same order of magnitude as the nonlinearity strength. Representative examples of VSs with $m=2$ and $4$ are shown in Fig.~\ref{fig2}. They exhibit a ring-shaped shape, similar to that of the potential in Fig.~\ref{fig1}(a). Yet, the amplitude and thickness of the stable vortices vary as functions of propagation constant $b$, while their radius remains, naturally, close to radius $r_{0}$ of the trapping potential, unlike the strongly expanding sable VSs in the free space \cite{Pego,michinel2004square}.

\begin{figure}[tbph]
\centering
\includegraphics[width=0.47\textwidth]{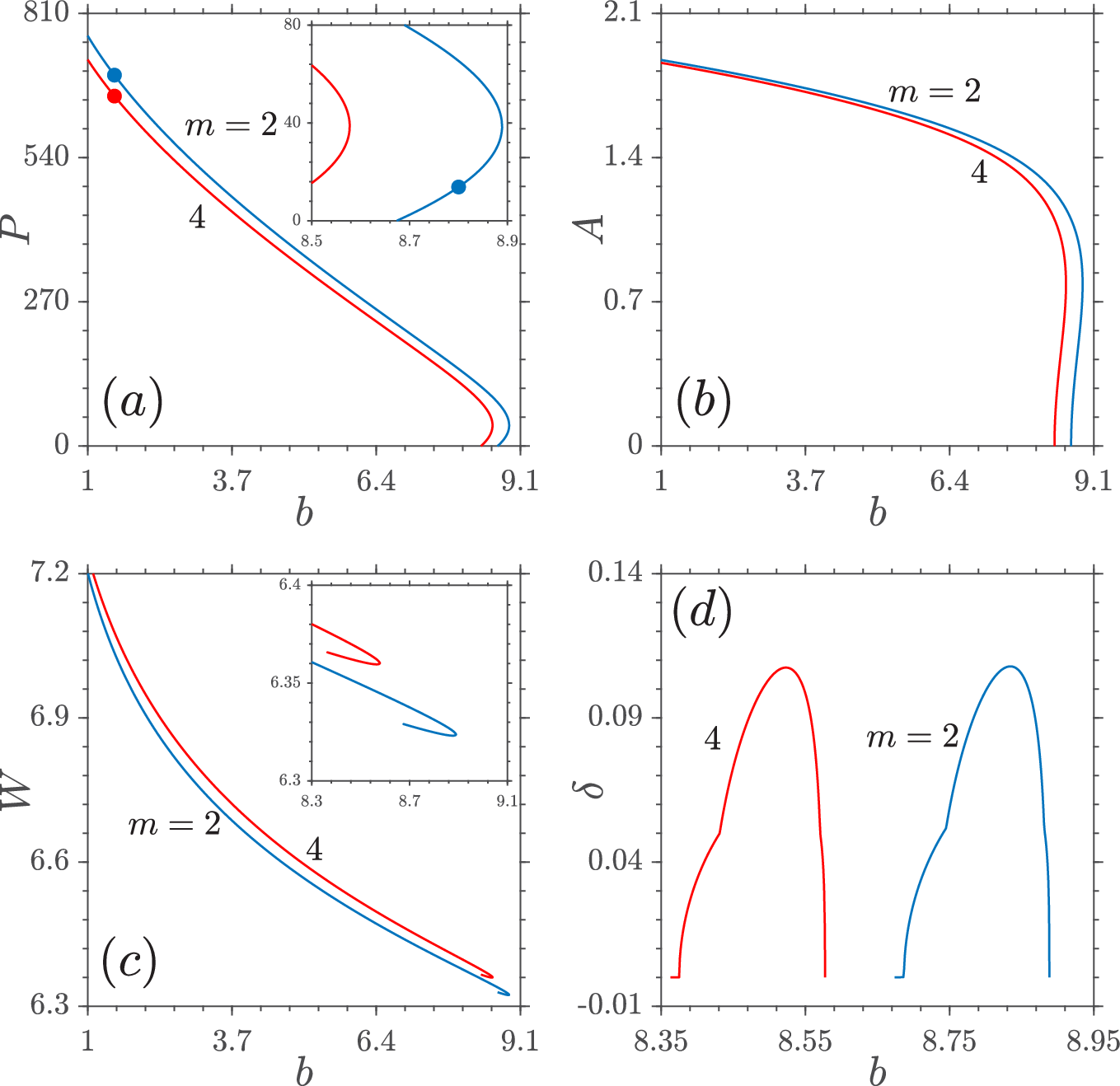}\vskip-0pc
\caption{(a) Power $P$ vs. propagation constant $b$ for VSs with $m=2$ and $4
$. (b) and (c): VS amplitudes and widths for $m=2$ and $4$ vs. $b$. Insets
in (a,c): zoom-in taken close to the cutoff (turning) point $b_{\text{cut}}$%
. (d) The instability growth rate $\protect\delta $ vs. $b$ for the lower VS
branches. In all panels, the potential's depth is $p=10$.}
\label{fig3}
\end{figure}

Unlike the free-space vortices, the power $P$ of the VSs considered here is
a nonmonotonous function of $b$, see Fig.~\ref{fig3}(a). For the VSs
originating from the linear modes at $b=8.674$ ($m=2$) and $b=8.362$ ($m=4$%
), $P$ increases with the growth of $b$, as long as the cubic term is the
dominant one. The VS's thickness and amplitude $|\psi |_{\text{max}}$
increase simultaneously. When $|\psi |_{\text{max}}$ exceeds $1$ (in the
scaled notation adopted here), the quintic term becomes dominant, slowing
the growth of $|\psi |_{\text{max}}$ [Fig.~\ref{fig3}(b)] and accelerating
the growth of the soliton's width, $W\equiv \sqrt{\int \int
(x^{2}+y^{2})\psi ^{2}\text{d}x\text{d}y/P}$ [Fig.~\ref{fig3}(c)]. The
transition from the focusing to defocusing nonlinearity, along with the
action of the external trapping potential, prevents the existence of VSs
with $b>b_{\text{cut}}$ and gives rise to another branch of the VSs with the
same $m$ but the opposite sign of the slope, $dP/db<0$. The two branches
merge at the cutoff point $b=b_{\text{cut}}$ [Figs.~\ref{fig3}(a,b)].

At fixed $b$, the power of the VS with $m=2$ is slightly higher than that
for $m=4$. The difference between the amplitudes of the upper-branch VSs
with $m=2$ and $4$ becomes very small as $b$ decreases [Fig.~\ref{fig3}(b)].
The difference between the width of the upper-branch VSs with $m=2$ and $4$
is also small, as seen in Fig.~\ref{fig3}(c). For example, at $b=1.5$, $%
\left\vert \left( |\psi |_{\text{max}}\right) _{\text{m=2}}-\left( |\psi |_{%
\text{max}}\right) _{\text{m=4}}\right\vert /\left( |\psi |_{\text{max}%
}\right) _{\text{m=2}}=0.0061$ and $\left\vert W_{\text{m=2}}-W_{\text{m=4}%
}\right\vert /W_{\text{m=2}}=0.0077$.

The main result of this work is that the linear-stability analysis,
performed on the basis of Eqs. (\ref{Eq4}), demonstrates that the ring
potential helps to maintain stable VSs with $m>2$, %
including large values of $m$. For small amplitudes $|\psi
|_{\text{max}}$, one may expect that the focusing nonlinearity dominated in
the lower-branch VSs, bifurcating from the linear vortex modes, makes the
vortices unstable. Indeed, for VSs with $m=2$ and $4$, there is only a very
narrow stability region near the linear limit, corresponding to $%
P\rightarrow 0$ [Fig.~\ref{fig3}(d)]. However, for the upper-branch VSs the
negative sign of $dP/db$ provides a necessary stability condition known as
the anti-Vakhitov-Kolokolov (anti-VK) stability criterion for solitons in
systems dominated by self-defocusing nonlinearities \cite{PhysRevA.81.013624}. Although, generally speaking, the anti-VK criterion is not sufficient for
the stability, in the present case the computation of the eigenvalue
spectrum produced by Eqs.~(\ref{Eq4}) demonstrates that the upper branch is
indeed entirely stable.

Note that the stable upper-branch vortices are narrow, and the
corresponding power is not extremely high [Figs.~\ref{fig3}(a) and (c)]. These features are in sharp contrast to the VSs in the uniform CQ medium \cite{michinel2004square}, where VSs with $m>2$ are stable only in very
narrow intervals of $b$, being extremely broad and carrying very high powers.

\begin{figure}[tbph]
\centering
\includegraphics[width=0.47\textwidth]{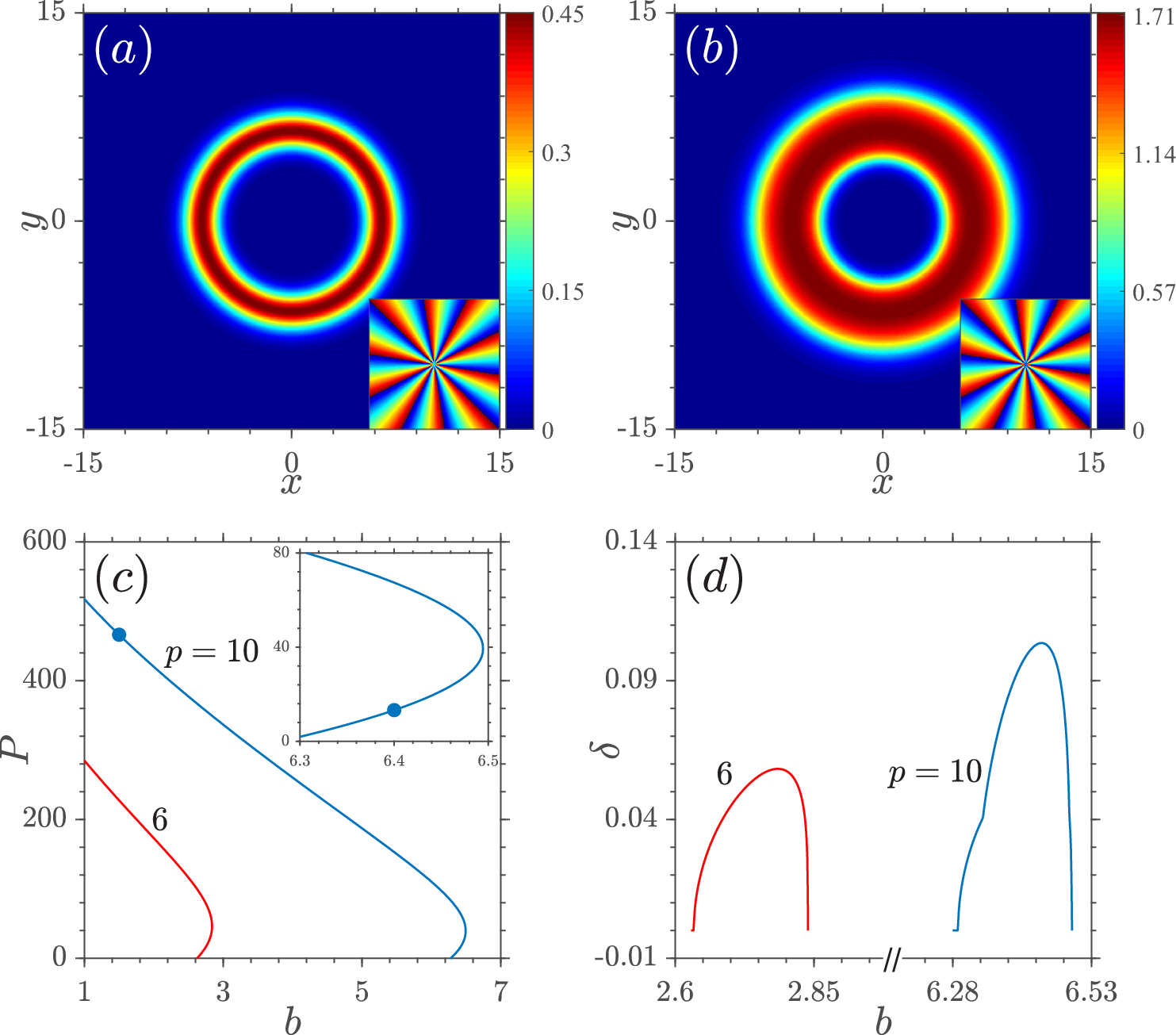}
\caption{(a) The map of $|\protect\psi (x,y)|$ in an unstable VS, with $m=10$
and $b=6.4$, belonging to the lower branch. (b) The map for a stable VS with
$m=10$ and $b=1.5$, belonging to the upper branch. These VSs correspond to
dots in (c), i.e., to $p=10$. Insets in (a) and (b) show the corresponding
phase patterns. (c) Power $P$ vs. $b$ for VSs with $m=10$, in the potential (%
\protect\ref{V}) with $p=10$ and $6$, respectively. Inset: zoom-in of $P(b)$
near the turning point, $b=b_{\text{cut}}$, for $p=10$. (d) The instability
growth rate $\protect\delta $ versus $b$ for the lower-branch VSs with $m=10$
in the potential with $p=10$ and $6$.}
\label{fig4}
\end{figure}

Addressing the challenging issue of stable VSs with high topological
charges, we have produced the solutions for up to $m=12$, see examples for $%
m=10$ in Fig. \ref{fig4}. Like their counterparts with lower values of $m$,
they feature shapes of \textquotedblleft light pipes", confined to the
narrow ring defined by the trapping potential (\ref{V}), although their
width and amplitude are somewhat larger, see Fig.~\ref{fig4}(b). In
accordance with what is stated above, the upper-branch VSs with $m=10$ are
stable in their entire existence domain, while their lower-branch
counterparts are unstable almost everywhere.

While examples of stable VSs are presented here for $m=2,4,8,10$ and $12$,
the same results hold as well for odd $m$ (at least, up to $m=11$). On the
other hand, we have checked that, if the defocusing quintic term is removed
in Eq. (\ref{Eq1}), all VSs generated by the remaining cubic NLSE with the
ring potential are completely unstable.
Thus, the inclusion of the quintic defocusing is crucially important for the
VS stability.

\begin{figure}[tbph]
\centering
\includegraphics[width=0.5\textwidth]{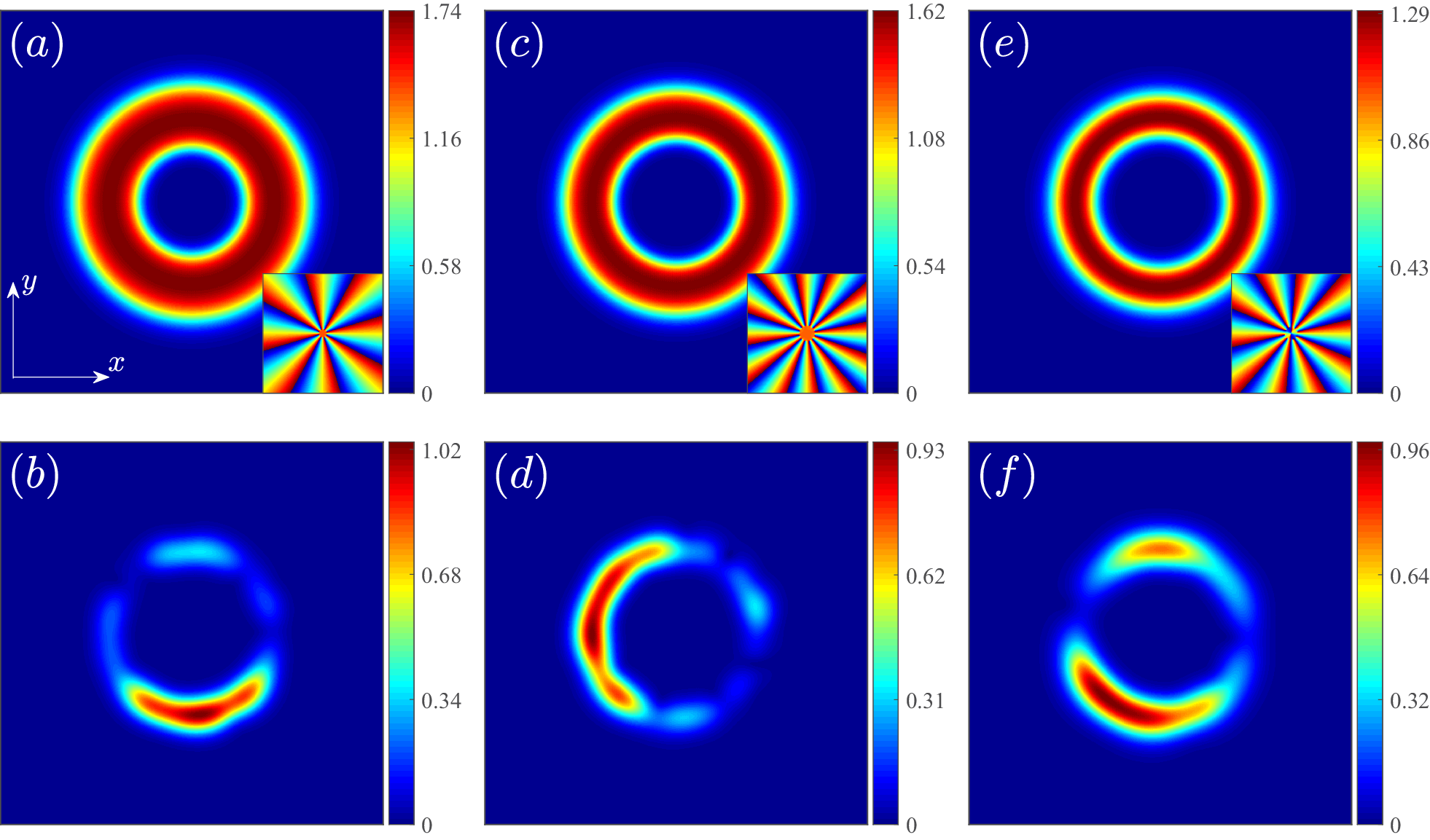}
\caption{The evolution of stable upper-branch (a,c,e) and unstable
lower-branch (b,d,f) VSs with topological charges $m=8$ in (a,b), $12$ in
(c,d), and $10$ in (e,f). The propagation constant is $b=2$ in (a,c,e), $7.3$
in (b), $5.4$ in (d), and $2.8$ in (f). The potential depth is $p=10$ in
(a-d) and $6$ in (e,f). Insets in (a,c,e) display the corresponding phase
patterns. The propagation distance is $z=2000$ and $(x,y)\in \lbrack -15,15]$
in all panels.}
\label{fig5}
\end{figure}

To verify the linear-stability predictions, we have performed systematic
simulations of the perturbed evolution by means of the split-step method
with the absorbing boundary conditions. To this end, the perturbed input was
taken as $\Psi (x,y,z=0)=\psi (x,y)[1+\rho (x,y)]$, where $\psi (x,y)$ is a
stationary VS, and $\rho (x,y)$ is random perturbation with variance $\sigma
_{\text{noise}}=0.01$. Typical results are displayed in Fig.~\ref{fig5}. The
upper-branch stable VSs preserve their structure in the course of very long
propagation, see Figs.~\ref{fig5}(a,c,e). On the other hand, unstable
lower-branch VSs break into one or more arcuate fragments [Figs.~\ref{fig5}%
(b,d,f)]. The breakup distance is $z\simeq 500$ and $300$ in Figs.~\ref{fig5}%
(b) and (d), respectively. Subsequently, the slightly varying fragments,
which keep the angular momentum of the initial VS, circulate along the ring
potential. Due to the compactness of the fragments, their amplitudes are
higher than $|\psi |_{\text{max}}$ of the input.

Finally, we discuss possibilities of the experimental creation of the
predicted VSs. In particular, polydiacetylene paratoluene sulfonate (PTS)
exhibits the CQ nonlinear RI, $\delta n=n_{2}I-n_{4}I^{2}$, as a function of
optical intensity $I$ \cite{Lawrence:98}. At wavelength $\lambda =1.6\mu$m, the nonlinear indices are $n_{2}=2.2\times 10^{-3}\text{cm}^{2}/\text{GW}$ and $n_{4}=0.8\times 10^{-3}\text{ cm}^{4}/\text{GW}^{2}$. The critical intensity at which $\delta n$ vanishes is $I_{0}=n_{2}/n_{4}=2.75\text{ GW/}\text{cm}^{2}$. Thus, the medium exhibits
the self-focusing and defocusing at $I<0.5I_{0}$ and $I>0.5I_{0}$,
respectively. The necessary ring-shaped profile of the linear
refractive-index may be fabricated, as mentioned above, by means of the
technology elaborated for production of composite fibers \cite{Payne}. %

To summarize, we have found that the effective trapping potential in the
form of a narrow ring, combined with the CQ nonlinearity, supports two
branches of VS (vortex-solitons) families with high values of topological
charge $m$. The radius and thickness of VSs are determined by the
ring-shaped potential. While the lower-branch VSs are almost completely
unstable, the upper branch is entirely stable, in sharp contrast to the VSs
trapped in the harmonic-oscillator potential \cite{LIU2023113422}, whose
stability region quickly shrinks with the growth of $m$. At least up to $m=12$, the stability of all VSs fully obeys the anti-VK criterion. We thus put
forward an experimentally relevant option for the creation of stable
higher-charge VSs, which was not proposed previously. The results are
relevant not only for optics but also for BEC, e.g., quantum droplets
trapped in ring-shaped optical potentials.

\vskip0.5pc \noindent\textbf{Funding.} Natural Science Basic Research Plan
in Shaanxi Province of China (2022JZ-02); Israel Science Foundation, grant
No. 1695/22.

\vskip0.5pc \noindent\textbf{Disclosures.} The authors declare no conflicts
of interest.

\vskip0.5pc \noindent\textbf{Data availability.} Data underlying the results
presented in this paper are not publicly available at this time but may be
obtained from the authors upon reasonable request.




\newpage { {\textbf{References with titles}} }

\end{document}